\documentclass[prb, preprintnumbers, amsmath, amssymb, superscriptaddress, twocolumn]{revtex4}
\usepackage{graphicx, color, dcolumn, bm, natbib, amssymb, amsmath}
\usepackage[final=true]{hyperref}

\newcommand{\m}{\mathbf}

\newcommand{\be}{\begin{eqnarray}}
\newcommand{\ee}{\end{eqnarray}}

\begin{document}

\title{Transition between long and short-range helical orders in mixed B20 cubic helimagnets}

\author{Oleg I. Utesov}
\email{utiosov@gmail.com}

\affiliation{Saint Petersburg State University, Ulyanovskaya 1, St. Petersburg 198504, Russia}
\affiliation{Petersburg Nuclear Physics Institute, Gatchina 188300, Russia}
\affiliation{St. Petersburg School of Physics, Mathematics, and Computer Science, HSE University, St. Petersburg 190008, Russia}

\begin{abstract}

A model of B20 helimagnet with strong (frustrating) defect bonds is considered. Single defect provides dipolar field-like distortion of the helical magnetic ordering. At finite defect bond concentrations, the spiral vector acquires correction to its value in the pure system. Moreover, it is shown that the strong defects of sufficient but small concentration are able to destroy long-range order driving the system to the phase with short-range helical ordering. It is argued that the developed approach allows describing main experimental observations in  Mn$_{1-x}$Fe$_x$Si and Mn$_{1-x}$Co$_x$Si compounds.

\end{abstract}
\maketitle

\section{Introduction}

In crystals without the center of inversion antisymmetric Dzyaloshinskii-Moriya interaction (DMI)~\cite{dzyal1958, moriya1960} causes instability of ferromagnetic ordering. As a result, one can observe helical~\cite{dzyal1964} and/or more complex noncollinear spin structures (e.g., skyrmions~\cite{bogdanov1989}). In this context, an important role is played by cubic helimagnets with B20 structure, e.g., MnSi and FeGe (see, e.g., Refs.~\cite{lundgren1970, ishikawa1976, Bak, nakanishi1980}). The recent upsurge of interest in this type of compound was partially stimulated by the first skyrmion lattice experimental observation~\cite{muhlbauer} and by the possibility of creating so-called racetrack memory using skyrmions~\cite{fert2013}.

Mixed B20 compounds, e.g., Mn$_{1-x}$Fe$_x$Ge, Mn$_{1-x}$Fe$_x$Si, Fe$_{1-x}$Co$_x$Si, etc., possess additional degree of freedom $x$, which opens up a possibility for the system parameters and properties fine tuning. For instance, in Mn$_{1-x}$Fe$_x$Ge the spiral vector gradually changes upon $x$ lowering from $x=1$ and becomes zero at $x_c \approx 0.75$~\cite{grig2013}. Moreover, in the vicinity of $x_c$ the system changes its magnetic chirality~\cite{grig2013} because effective DMI constant $D$ goes through zero~\cite{kikuchi2016}. In contrast, Mn$_{1-x}$Fe$_x$Si and Mn$_{1-x}$Co$_x$Si demonstrate large variation of the spiral vector at $x \ll 1$ accompanied by a significant suppression of $T_c$ and spin-wave stiffness~\cite{grig2009,bauer2010, grig2011, bannenberg2018, grig2018,kindervater2020}. Furthermore, there is a transition between the long-range order and the short-range one at the quantum critical point $x_c \approx 0.1$~\cite{bauer2010, grig2011, demishev2014}.

Previously, we have developed a disordered lattice approach for helimagnets~\cite{utesov2015,utesov2019}. It was shown that defect bonds (characterized by different from the pure ones DMI and/or exchange coupling) lead to the spiral vector variation as well as helical ordering distortions. However, these effects are rather small (but observable) at small defects concentrations being proportional to the defects concentration.

\begin{figure}
  \centering
  \includegraphics[width=8cm]{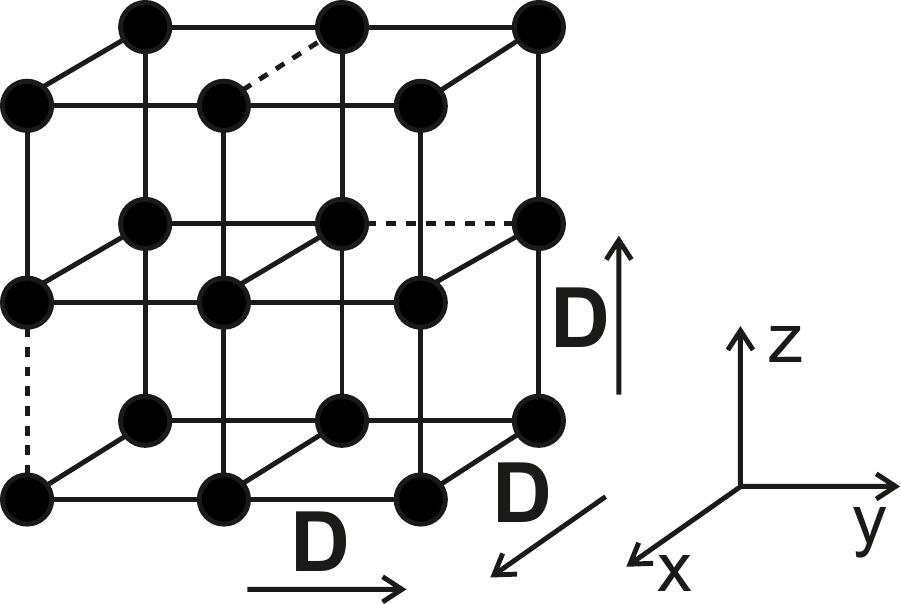}\\
  \caption{In the considered model magnetic ions are placed on the sites of a simple cubic lattice. They interact via Dzyaloshinskii-Moriya interaction and conventional exchange coupling, but some bonds are defect ones (dashed lines) with different values of the parameters $D$ and $J$ from the regular bonds (solid lines). }\label{fig1}
\end{figure}

In the present paper, we modify the approach of Refs.~\cite{utesov2015,utesov2019} to the case of ``strong'' defect bonds. Using the localized nature of magnetic moments of B20 helimagnets~\cite{demishev2011} and hints of antiferromagnetism enhancement in  Mn$_{1-x}$Fe$_x$Si upon $x$ growth~\cite{glushkov2015, bannenberg2018}, we formulate a model of disordered B20 ``Bak-Jensen'' helimagnet with strong (frustrating) antiferromagnetic bonds randomly distributed over the sample with a certain concentration $c$ (see Fig.~\ref{fig1}). Such bonds were previously shown to provide dipolar-like distortions of the magnetic order in collinear magnets~\cite{aharony, korenblit} and can even destroy long-range order~\cite{cherepanov1999}. We show that strong defects in our model still provide dipolar-like perturbation of the helimagnetic ordering; however, in contrast to Refs.~\cite{utesov2015,utesov2019}, the corresponding ``charge'' should be found by solving a nonlinear equation of self-consistency. This result is also supported by our numerics. We argue that the proposed here (rather simple) model allows describing on semi-quantitative level main experimental observations in  Mn$_{1-x}$Fe$_x$Si and Mn$_{1-x}$Co$_x$Si in a wide range of concentration $x$, which is associated with the theoretical parameter $c$.

The rest of the paper is organized as follows. In Sec.~\ref{SForm} we introduce the model and formulate the problem. The theory for a single defect bond is presented in Sec.~\ref{SSstrong}. It is used for systems with finite defect bond concentrations addressed in Sec.~\ref{SFinite}, which also contains an application to real compounds. In Sec.~\ref{SConc} we present our conclusions. In Appendix~\ref{AppendixA} we compare our analytical predictions for a single defect problem with Monte-Carlo simulations using particular parameters set.

\section{Formulation of the problem}

\label{SForm}

As a starting point of our analysis, we consider Hamiltonian of the following form~\cite{Bak,maleyev} on a simple cubic lattice (see Fig.~\ref{fig1}):
\begin{eqnarray} \label{ham1}
  \mathcal{H}_0 &=& \mathcal{H}_{ex} + \mathcal{H}_{dm}, \\
  \mathcal{H}_{ex} &=& -\frac{1}{2} \sum_{\mathbf{R},\mathbf{R}^\prime} J_{\mathbf{R},\mathbf{R}^\prime} \mathbf{S}_\mathbf{R} \cdot \mathbf{S}_{\mathbf{R}^\prime},  \\
  \mathcal{H}_{dm} &=& -\frac{1}{2} \sum_{\mathbf{R},\mathbf{R}^\prime} \mathbf{D}_{\mathbf{R},\mathbf{R}^\prime} \cdot \left[ \mathbf{S}_\mathbf{R} \times \mathbf{S}_{\mathbf{R}^\prime} \right],
\end{eqnarray}
Note that we neglect small anisotropic interactions (anisotropic exchange and cubic anisotropy~\cite{Bak,maleyev}) in the calculations below. However, we bear in mind that they are important for the determination of the spiral vector direction, which can be either along the cubic edge or spatial diagonal (these two cases should be treated separately). Next, for definiteness, we consider only nearest-neighbors interactions with constants $J$ and $D$ for exchange interaction and DM, respectively. In such a case, the spiral vector reads
\begin{equation}\label{qsp}
  q = \frac{D}{J} \ll 1,
\end{equation}
where we set lattice parameter $a=1$. Importantly, for pure MnSi it means that the angle between neighboring spins is rather small, being $\sim 0.1$ in radians.

In our two previous papers~\cite{utesov2015,utesov2019}, a theory of mixed noncentrosymmetric helimagnets with the bond disorder was developed. The main yield was that if the Hamiltonian~\eqref{ham1} is perturbed by
\begin{equation}\label{pert1}
  \mathcal{V} = - \sum_i \left( U_{ex} \mathbf{S}_{\mathbf{R}_i} \cdot \mathbf{S}_{\mathbf{R}^\prime_i} + U_{dm} \mathbf{e}_{\mathbf{R}_i,\mathbf{R}^\prime_i} \cdot \left[ \mathbf{S}_{\mathbf{R}_i} \times \mathbf{S}_{\mathbf{R}^\prime_i} \right] \right),
\end{equation}
index $i$ enumerates the defect bonds, then the spiral vector $q$~\eqref{qsp} acquires finite linear in defects concentration $c$ correction $\delta q$, provided that $c \ll 1$. We define $c$ so that $c/3$  is the probability for each particular bond to be the defect one. In more detail, it was shown that
\begin{equation}\label{corr1}
  \delta q \propto c \frac{U_{dm} - U_{ex} q }{J(1 + U_{ex}/3 J)}.
\end{equation}
Evidently, $\delta q \ll q$ if the denominator is not very small. Moreover, it can be shown that $(U_{dm} - U_{ex} q)/J(1 + U_{ex}/3 J)$ is proportional to additional relative rotation of the spins on the defect bond.

Remarkably, in the Mn$_{1-x}$Fe$_x$Si compound, a linear dependence of the spiral vector on $x$ was observed~\cite{grig2018, kindervater2020}. However, for $x \approx 0.1$ the variation of the spiral vector $\delta q \approx q$. In the framework of the developed approach ~\cite{utesov2015,utesov2019}, it is natural to connect $x$ and the theoretical model parameter $c$. It means that the denominator in Eq.~\eqref{corr1} should be small to compensate for the smallness of $c$. So, the additional rotation of the spins on defect bonds is large, being of the order of unity. The approach of Refs.~\cite{utesov2015,utesov2019} was developed under the assumption of small additional distortions of the spiral ordering due to the imperfect bonds, hence it is not applicable for a description of ``strong'' defects. However, below we argue that the model with strong defects can be used for Mn$_{1-x}$Fe$_x$Si properties description. So, the approach of Refs.~\cite{utesov2015,utesov2019} should be properly adapted.

\section{Single strong defect}
\label{SSstrong}

We start our theoretical discussion with the one-impurity problem. For definiteness, we first assume that the spiral vector of the unperturbed system is oriented along the $z$ axis and that the defect bond connects sites $\mathbf{R}_1=(0,0,1)$ and $\mathbf{R}_0=(0,0,0)$. Then, the perturbation~\eqref{pert1} reads
 \begin{equation}\label{pert2}
  \mathcal{V} = - U_{ex} \mathbf{S}_{(0,0,1)} \cdot \mathbf{S}_{(0,0,0)} - U_{dm} \mathbf{e}_{z} \cdot \left[ \mathbf{S}_{(0,0,1)} \times \mathbf{S}_{(0,0,0)} \right].
\end{equation}
Each of the spins (which are assumed to lie in the $xy$ plane) is described by the polar angle $\varphi_{\mathbf{R}}$. Then, the system classical energy is a function of the angles, explicitly
\begin{eqnarray} \label{en1}
  \frac{E[\varphi]}{S^2} &=& - J \sum_{\langle \mathbf{R},\mathbf{R}^\prime \rangle} \cos{(\varphi_{\mathbf{R}}- \varphi_{\mathbf{R}^\prime})} \nonumber \\ &&- D \sum_{\mathbf{R}} \sin{(\varphi_{\mathbf{R}+\mathbf{e}_z}- \varphi_{\mathbf{R}})} \\ &&- U_{ex} \cos{(\varphi_{\mathbf{R}_1}- \varphi_{\mathbf{R}_0})} - U_{dm}  \sin{(\varphi_{\mathbf{R}_1}- \varphi_{\mathbf{R}_0})} \nonumber,
\end{eqnarray}
where the first summation includes all neighboring pairs of spins.

Spin ordering with minimal energy has first derivatives on $\varphi$ variables equal to zero, which provides the following equation for all sites $\mathbf{R}$:
\begin{eqnarray} \label{sys1}
  && J \sum_{\mathbf{R}^\prime} \sin{(\varphi_{\mathbf{R}}- \varphi_{\mathbf{R}^\prime})} - \nonumber \\ && D \left[ \cos{(\varphi_{\mathbf{R}}- \varphi_{\mathbf{R}-\mathbf{e}_z})} - \cos{(\varphi_{\mathbf{R}+\mathbf{e}_z}- \varphi_{\mathbf{R}})}\right] \nonumber \\
  &&+  (\delta_{\mathbf{R},\mathbf{R}_1} - \delta_{\mathbf{R},\mathbf{R}_0}) [U_{ex} \sin{(\varphi_{\mathbf{R}_1}- \varphi_{\mathbf{R}_0})} \\ &&- U_{dm}\cos{(\varphi_{\mathbf{R}_1}- \varphi_{\mathbf{R}_0})}]=0. \nonumber
\end{eqnarray}
This system of equations is nonlinear and can not be solved directly. However, according to our Monte-Carlo simulations (see Appendix~\ref{AppendixA} for some details), all the angles differences on nearest neighbors are small enough, and corresponding sine and cosine functions can be expanded in their powers, except for $\varphi_{\mathbf{R}_1}- \varphi_{\mathbf{R}_0} \equiv \delta \varphi$. Then, for all $\mathbf{R} \neq \mathbf{R}_1,\mathbf{R}_0$ Eq.~\eqref{sys1} approximately reads
\begin{equation}\label{sys2}
  J \sum_{\nu=x,y,z} (2 \varphi_\mathbf{R} - \varphi_{\mathbf{R} +\mathbf{e}_\nu} - \varphi_{\mathbf{R} - \mathbf{e}_\nu})= 0,
\end{equation}
which is (cf. Ref.~\cite{utesov2015}) equivalent to a lattice version of the Laplace equation, $\Delta \varphi = 0$. In contrast, for sites $\mathbf{R}_1$ and $\mathbf{R}_0$ after some transformations one obtains
\begin{equation}\label{sys3}
  - J \Delta \varphi = 4 \pi Q (\delta_{\mathbf{R},\mathbf{R}_1} - \delta_{\mathbf{R},\mathbf{R}_0}),
\end{equation}
where
\begin{eqnarray}\label{Q1}
  Q &=& \frac{1}{4 \pi} \bigl[ J(\delta \varphi - \sin{\delta \varphi}) - U_{ex} \sin{\delta \varphi} \\ &&- D(1-\cos{\delta \varphi}) + U_{dm} \cos{\delta \varphi}\bigr]. \nonumber
\end{eqnarray}
One can see that, as in the previous papers~\cite{utesov2015,utesov2019}, spin angles satisfy Poisson's equation for electric dipole. In this case, we write its solution for sites not connected by the defect bond as follows:
\begin{equation}\label{Dip1}
  \varphi_{\mathbf{R}} = \frac{Q}{J}\left( \frac{1}{|\mathbf{R}-\mathbf{R}_1|} - \frac{1}{|\mathbf{R}-\mathbf{R}_0|} \right) + q (R_z-1/2).
\end{equation}
The last term here is the solution of homogeneous equation for the system without the defect. Finally, we should find $\delta \varphi$ and, correspondingly, the ``charge'' $Q$. It can be done self-consistently, since Eq.~\eqref{Dip1} works quite well quantitatively even for sites neighboring to $\mathbf{R}_1$ and $\mathbf{R}_0$. Using Eq.~\eqref{Dip1} we find that $\varphi_{(1,0,1)} = -\varphi_{(1,0,0)} = (1-1/\sqrt{2})Q/J + q/2$, $\varphi_{(0,0,2)} = -\varphi_{(0,0,-1)} = Q / 2 J + 3q/2$; other important angles can be found using symmetry arguments. Considering two equations of the system~\eqref{sys3} for $\mathbf{R}_1$ and $\mathbf{R}_0$, and subtracting one from another, we derive the condition of self-consistency, which should be solved to find $\delta \varphi$. It reads
\begin{equation}\label{Q2}
  \delta \varphi \approx \frac{D}{J} + \frac{4 \pi Q}{3 J}.
\end{equation}
If $\delta \varphi \ll 1$, this equation is equivalent to linear one which was solved in Refs.~\cite{utesov2015, utesov2019}. In the opposite case, $\varphi \sim 1$ and one can neglect DMI-induced terms and obtain
\begin{equation}\label{Q3}
  \frac{2}{3} \delta \varphi = - \frac{ J + U_{ex}}{3 J} \sin{\delta \varphi}.
\end{equation}
This equation has nontrivial solutions only if $U_{ex} < - 3 J$ (strong antiferromagnetic exchange on the defect bond), which provides local frustration of the exchange interaction (see also Refs.~\cite{aharony,korenblit}). Remarkably, for the same parameters our previous approach fails (see Eq.~\eqref{corr1}, and Ref.~\cite{utesov2015} for the details). Note, that Eq.~\eqref{Q3} has $\delta \varphi \leftrightarrow - \delta \varphi$ symmetry, which appears after we neglect all small DMI contributions. In reality, $\delta \varphi$ has the same sign as $D$ due to the energy reasons provided $|U_{dm}| \sim |D|$.

Using the same formalism, we can describe an influence of the ``horizontal'' defect bond on the spin ordering, e.g., between the sites with $\mathbf{R}_1=(1,0,0)$ and $\mathbf{R}_0=(0,0,0)$. Similar calculations show that in this case, Eq.~\eqref{Q3} determines $\delta \varphi$ without any additions from DMI (e.g., without the first term on the right-hand side of Eq.~\eqref{Q2}), and the spin ordering is defined by Eq.~\eqref{Dip1}. Importantly, in this case, nonzero $Q$ also requires $U_{ex} < - 3 J$. Moreover, this defect has $\delta \varphi \leftrightarrow - \delta \varphi$ symmetry, and at finite defects concentration one will have both types of dipoles with equal probability. Evidently, such defects can not affect the spiral vector. Nevertheless, they provide distortions of the spin ordering.

When considering the spiral vector oriented along with one of the cubic spatial diagonals, e.g., $\m{q} = (q,q,q)/\sqrt{3}$, we arrive at the same conclusions as previously. The counterpart of Eq.\eqref{Dip1} reads
\begin{equation}\label{Dip2}
  \varphi_{\mathbf{R}} = \frac{Q}{J}\left( \frac{1}{|\mathbf{R}-\mathbf{R}_1|} - \frac{1}{|\mathbf{R}-\mathbf{R}_0|} \right) + q \frac{R_x + R_y +R_z}{\sqrt{3}},
\end{equation}
and simple calculations show that only minor change of Eq.~\eqref{Q2} is in order:
\begin{equation}\label{Q4}
  \delta \varphi \approx \frac{D}{\sqrt{3}J} + \frac{4 \pi Q}{3 J}.
\end{equation}
Finally, in the case of strong defects, DMI terms in this equation can be neglected, and one can see that Eq.~\eqref{Q3} stays intact.

\section{Finite defects concentrations. Application to $\textrm{Mn}_{1-x}\textrm{Fe}_x\textrm{Si}$}

\label{SFinite}


First, we address the problem of the system with finite concentration $c$ of the defect bonds. In this case, the electrostatic analogy can be further exploited (see also Ref.~\cite{utesov2015}), and the system is, on average, equivalent to a uniformly polarized one. Let, for definiteness, $\mathbf{q} \parallel \hat{z}$. Due to the defect bonds oriented along $z$-axis, we have the ``polarization'' (contributions from perpendicular bond directions cancel each other)
\begin{equation}\label{pol1}
  \mathbf{P} = \frac{c}{3} \frac{Q}{J} \mathbf{e}_z.
\end{equation}
Importantly, as in each media with dipolar fields, the sample shape can become crucial. However, in our case, the boundary condition (it can be derived from Eq.~\eqref{sys1}) reads $\partial \varphi/\partial \m{n} =0 $ ($\m{n}$ is a normal to the surface vector), and the ``electrical field'' is transverse at the boundary. So, its flow is ``confined'' inside the sample, and we can write the equation of Ref.~\cite{utesov2015}
\be
  \frac{\partial \varphi}{\partial z} = 4 \pi P = \frac{4 \pi c Q}{3 J},
\ee
which is equivalent to an additional rotation of each spin in the system, i.e., correction to the spiral vector
\begin{equation}\label{corr2}
  \delta q = c \frac{4 \pi Q}{3 J}.
\end{equation}
Moreover, finite defect bonds concentration leads to quenched transverse fluctuations of the ordered spin value, which can be written as
\begin{equation}\label{fluc1}
  \langle \delta S^2_\perp \rangle/S^2 = \langle \sin^2{\delta \varphi_\mathbf{0}} \rangle.
\end{equation}
Here $\delta \varphi_\mathbf{0}$ is the deviation of a certain spins angle (e.g., in the origin of coordinates) from its average value, and the averaging is taken over the disorder configurations. Eq.~\eqref{fluc1} can be simplified by virtue of $\delta \varphi_\mathbf{0}$ smallness in the majority of the disorder configurations when it is not connected with another site by the defect bond. So, we obtain
\begin{equation}\label{fluc2}
  \langle \delta S^2_\perp \rangle/S^2 \approx c \sum_\mathbf{R} \delta \varphi^2_\mathbf{R},
\end{equation}
with $\delta \varphi_\mathbf{R}$ given by the dipolar part of the solution~\eqref{Dip1}.

Note, this equation (as well as the one for the spiral vector correction~\eqref{corr2}) neglects the defects' ``interference'' effects. It is correct only if the distance between neighboring defect bonds is much larger than the lattice parameter, i.e., $c^{-1/3} \gg 1$. So, at $c \gtrsim 0.1$ theoretical predictions can be correct only semi-quantitatively.

The quenched transverse fluctuations at $c \ll 1$ were previously shown to diminish magnetic spiral Bragg peak and to provide diffusive background in elastic neutron scattering~\cite{utesov2015, utesov2019}. However, in lower dimensions they destroy long-range order in non-collinear~\cite{santanu2020} and collinear (see Refs.~\cite{aharony,korenblit,cherepanov1999}) magnets. This phenomenon can be addressed using a continuous version of Eq.~\eqref{fluc2},
\begin{equation}\label{fluc3}
  \langle \delta S^2_\perp \rangle/S^2 \approx c \int d^D \m{r} \delta \varphi^2(\mathbf{r}).
\end{equation}
The right hand side of this equation for dipolar field $\delta \varphi(\mathbf{r})$ diverges at large $\m{r}$ for $D \leq 2$ no matter how small $c$ is. However, in the 3D case, one needs strong defect to destroy the long-range order~\cite{cherepanov1999}. Quantitatively, the corresponding condition can be written as
\begin{equation}\label{fluc4}
  \langle \delta S^2_\perp \rangle/S^2 \sim 1.
\end{equation}

In order to address the Mn$_{1-x}$Fe$_x$Si properties, we for definiteness treat variable $c$ as an equivalent of $x$. The first property, which should be addressed is the previously mentioned  $\delta q \approx q$ at $x=0.1$~\cite{grig2018, kindervater2020} condition. For pure MnSi dimensionless $q=D/J\approx 0.2$. In the framework of the developed approach, we take $U_{ex} = -6 J$, which yields (see Eqs.~\eqref{Q2},~\eqref{Q3}, and~\eqref{corr2}) $Q \approx 0.5 J$ and $\delta q \approx 0.2$ for $c=0.1$. Note, that here and below we neglect small $U_{dm}$ corrections.

Next, according to the condition~\eqref{fluc4}, such $Q$ provides transition to the short-range order phase at some $c_{crit} \sim 0.1$. At $c> c_{crit}$ we can estimate the correlation length $c$-dependence using the following equation:
\be
  c_{crit} \int d^3 \m{r} \delta \varphi^2(\mathbf{r}) = c \int^\xi d^3 \m{r} \delta \varphi^2(\mathbf{r}),
\ee
where in the right-hand side we introduced the cut-off -- correlation length $\xi$. It yields ($\varphi(\mathbf{r}) \sim 1/r^2$ at large $r$)
 \begin{equation}\label{leng}
   \xi(c) \propto \frac{1}{c-c_{crit}}.
 \end{equation}
This result lies in agreement with the renormalization group ($D=2+\epsilon$-expansion) prediction of Ref.~\cite{cherepanov1999}. So, at $c>c_{crit}$ even at $T=0$ the correlation length is finite, which is a low-temperature analogue of the highly chiral fluctuating (HCF) phase visible at $T_{dm}>T>T_c$~\cite{grigoriev2005}. Importantly, this means that $T_c=0$ at $c>c_{crit}$ and at low temperatures $T \ll T_{dm}$ the system freezes in some short-range helical order. Note that finite correlation length should play an important role in the broadening of magnetic susceptibility peaks at $T_{DM}$, which was observed experimentally in Ref.~\cite{bannenberg2018}.

Moreover, in the present model average exchange constant $\langle J \rangle = J + c U_{ex}$ decreases with $c$ increasing due to antifferomagnetic ($U_{ex} + J < 0$) exchange on the defect bonds. It results in a decrease of the spin waves stiffness and $T_{dm}$ observed experimentally~\cite{grig2018}, both of these quantities being $\propto \langle J \rangle$. When \mbox{$J + c U_{ex} = 0$} one has $T_{dm}=0$ (for parameters considered here it gives $c \approx 0.17$), which means transition to the Griffiths-type~\cite{Griffiths1969} phase where only composition fluctuations-induced small helical ``islands'' are present (see, e.g., Ref.~\cite{glushkov2015}).

\section{Conclusions}

\label{SConc}

We discuss a model of B20 helimagnet with strong (frustrating) defect bonds. We show, that the familiar from previous papers dipolar-like solution of the single defect problem is also applicable in this case, but the charge is to be found from the non-linear equation of self-consistency. Moreover, at finite defects concentrations, significant variations of the spiral pitch can be observed even at $c \ll 1$. In the regime of moderate $c \gtrsim 0.1$ defects can destroy long-range order and provide the short-range one with finite spin correlation length.

We argue that the proposed model provides a semi-quantitative description of the set of important features observed experimentally in Mn$_{1-x}$Fe$_x$Si, namely: (i) large variation of the spiral vector value at small $x$, (ii) transition between long-range and short-range magnetic orderings, (iii) decrease of the spin-wave stiffness under $x$ growth, and (iv) subsequent disappearance of magnetically ordered phase at even larger $x$.  We point out that, without any doubt, presented here theory is a simple model description of the real physics. However, we believe that its qualitative predictions can be useful in the discussion of this material peculiar properties. Moreover, this model has been already successfully used for qualitative description of neutron experiments, see Ref.~\cite{grigoriev2021}. Finally, we note that the similar behavior of the related compound Mn$_{1-x}$Co$_x$Si~\cite{bauer2010, kindervater2020} can be also discussed using the developed approach.

\begin{acknowledgments}

We are grateful to A.V.\ Syromyatnikov for valuable discussions. This work is supported by the Russian Science Foundation (Grant No. 22-22-00028).

\end{acknowledgments}

\appendix

\section{Numerical modeling}
\label{AppendixA}

In order to verify our self-consistent approach for a single defect problem discussed in Subsec.~\ref{SSstrong}, we perform classical Monte-Carlo simulations at $T=0$. We analyzed three-dimensional lattices of sizes from $20\times20\times20$ to $50\times50\times50$ sites, the number of steps was $5 \times 10^6 - 2 \times 10^7$.

\begin{figure}
  \centering
  \includegraphics[width=6cm]{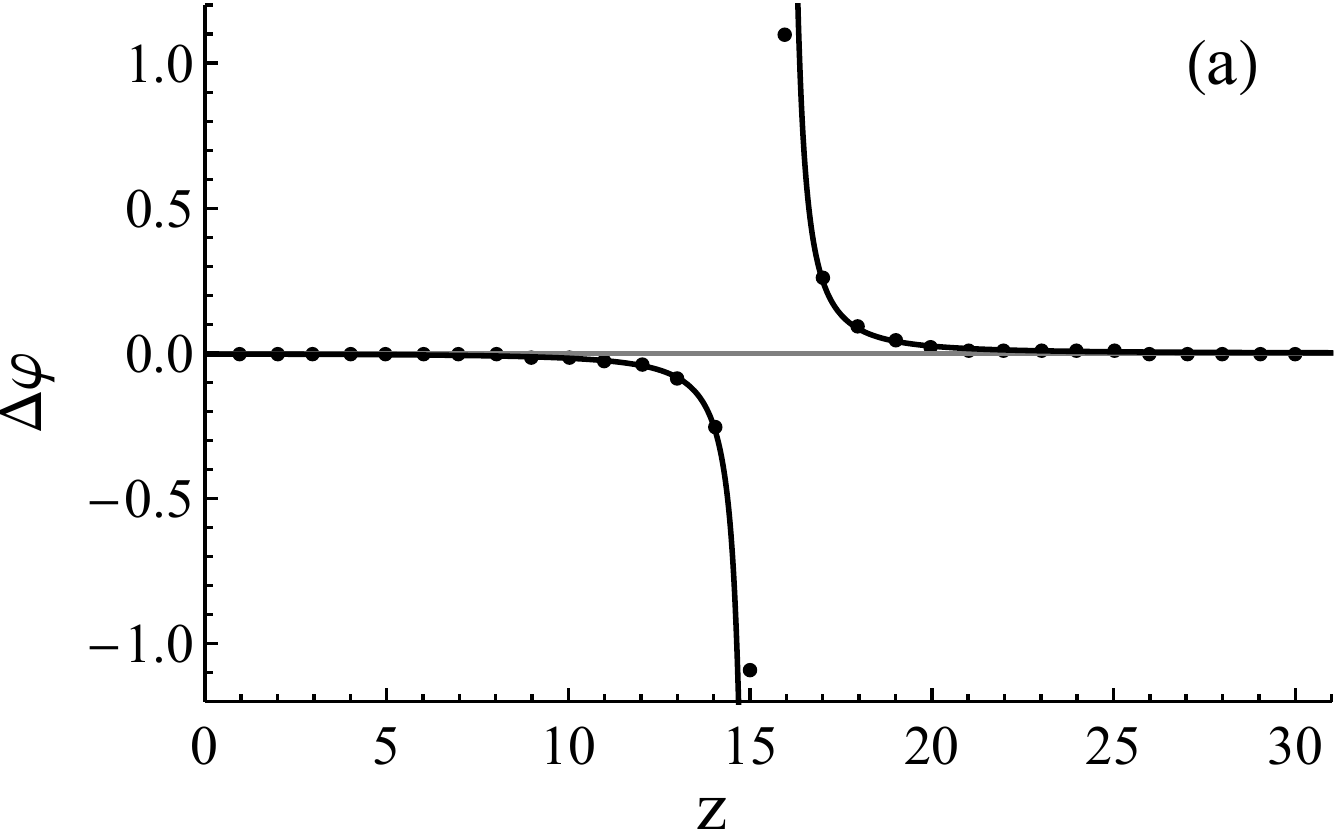}\\
  \includegraphics[width=6cm]{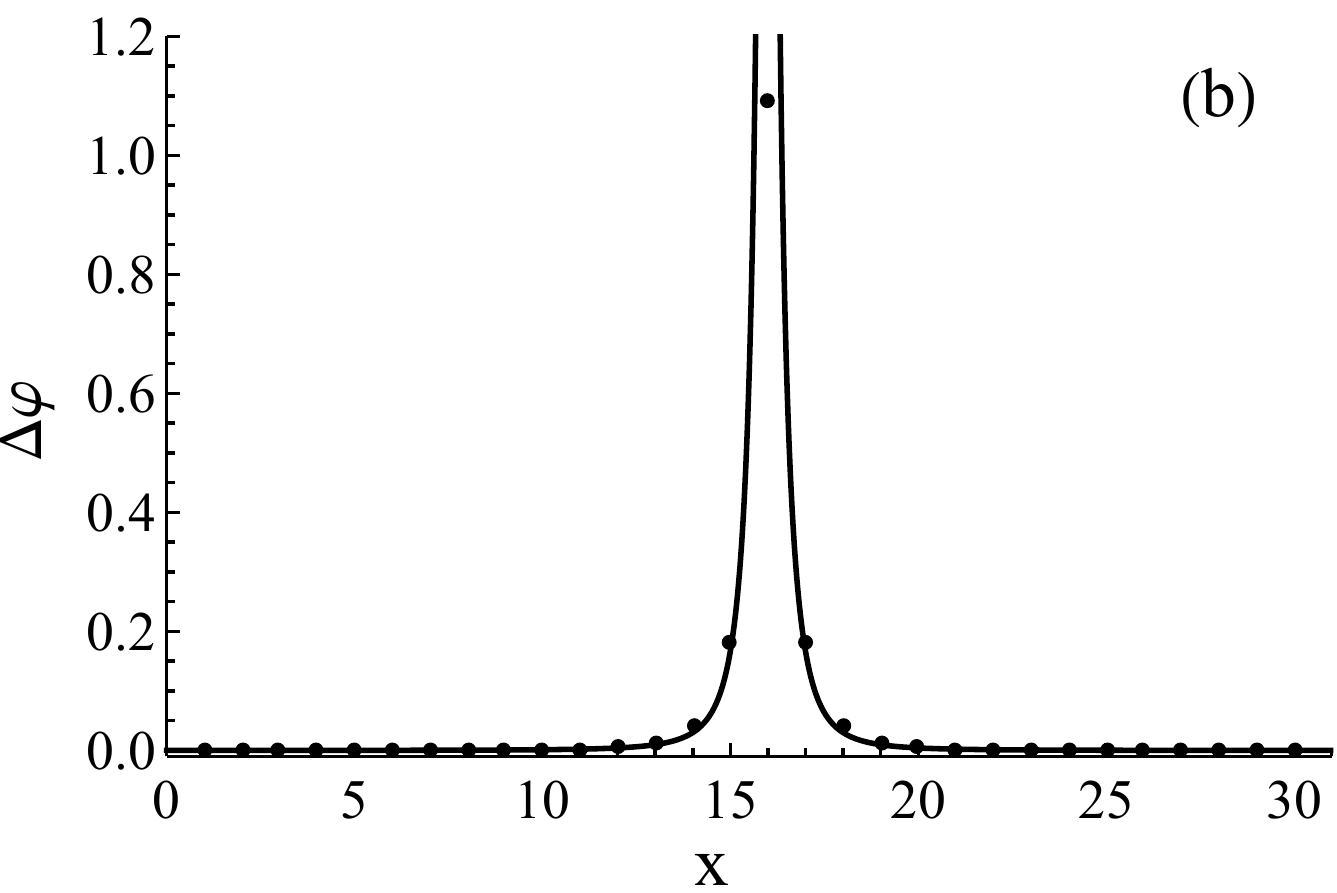}\\
  \caption{Additional rotations of spins induced by the strong defect oriented along the $z$ axis (which also is chosen to be a spiral vector direction) for the line including the defect bond (a) and in the perpendicular direction (b). Black dots are the result of numerical modeling discussed in Appendix~\ref{AppendixA}, while black curves illustrate the analytical self-consistent approach (see Sec.~\ref{SSstrong}). In both panels significant spiral ordering distortions near the defect bond are pronounced. }\label{figA}
\end{figure}

For instance, taking parameters set:
\begin{equation}\label{AppPar}
  J=1, \quad D=0.1, \quad U_{ex}=-6, \quad U_{dm}=-0.3,
\end{equation}
for defect oriented along the spiral vector we obtain numerically $\delta \varphi \approx 131^\circ$, whereas analytical Eq.~\eqref{Q2} yields $\delta \varphi \approx 126^\circ$. Then, the ``charge'' $Q \approx 0.5$ and we can compare numerical $\varphi_\mathbf{R}$ with the analytical one given by Eq.~\eqref{Dip1}. Corresponding results are shown in Fig.~\ref{figA}, where the homogeneous part (helical rotation) was subtracted, and the only additional defect-induced term is shown. There is a good agreement between the numerics and the analytical approach even for neighboring the defect bond sites. In the case of the defect bond perpendicular to the spiral vector, we observe almost the same physical picture.

\bibliography{bibliography}

\end{document}